\documentclass[aps,letterpaper,nofootinbib,preprint,showpacs,preprintnumbers,amsmath,amssymb]{revtex4}%
\usepackage{latexsym}%
\usepackage{hyperref}%

\def\oR{{\overline R}}
\def\oP{{\overline P}}
\def\otau{{\overline \tau}}
\def\tR{{\widetilde R}}
\def\cF{{\mathcal F}}
\def\cH{{\mathcal H}}

\def\cO{{\mathcal O}}
\def\cP{{\mathcal P}}
\def\cQ{{\mathcal Q}}
\def\cW{{\mathcal W}}
\def\bbH{{\mathbb H}}
\def\bbF{{\mathbb F}}
\def\hO{{\widehat O}}
\def\hP{{\widehat P}}

\newcommand{\beq}{\begin{equation}}
\newcommand{\beqn}{\begin{equation}\nonumber}
\newcommand{\eeq}{\end{equation}}
\newcommand{\bea}{\begin{eqnarray}}
\newcommand{\bean}{\begin{eqnarray}\nonumber}
\newcommand{\eea}{\end{eqnarray}}

\begin{document}

\begin{center}
{\bf {\Large Quantum Gravitational Collapse and}}\\
\smallskip
{\bf {\Large Hawking Radiation in 2+1 Dimensions}}

\bigskip
\bigskip

{{Cenalo Vaz$^{a}$\footnote{e-mail address: Cenalo.Vaz@UC.Edu},
Sashideep Gutti$^{b}$\footnote{e-mail address: sashideep@tifr.res.in},
Claus Kiefer$^{c}$\footnote{e-mail address: kiefer@thp.uni-koeln.de}, and
T. P. Singh$^{b}$\footnote{e-mail address: tpsingh@tifr.res.in}}}

\bigskip
\bigskip

\medskip

{\it $^{a}$RWC and Department of Physics, University of Cincinnati}\\
{\it Cincinnati, Ohio 45221-0011, USA}

\medskip

{\it $^{b}$Tata Institute of Fundamental Research,}\\
{\it Homi Bhabha Road, Mumbai 400 005, India}

\medskip

{\it $^{c}$Institut f\"ur Theoretische Physik, Universit\"at zu K\"oln}\\
{\it Z\"ulpicher Str. 77, 50937 K\"oln, Germany}

\end{center}

\medskip

\centerline{ABSTRACT}

\noindent
We develop the canonical theory of gravitational collapse in 2+1
dimensions with a
negative cosmological constant and obtain exact solutions of the Wheeler--DeWitt 
equation regularized on a lattice. We employ these solutions to derive the Hawking 
radiation from black holes formed in all models of dust collapse. We obtain an 
(approximate) Planck spectrum near the horizon characterized by the Hawking 
temperature $T_{\mathrm H}=\hbar\sqrt{G\Lambda M}/2\pi$,
 where $M$ is the mass of a black 
hole that is presumed to form at the center of the collapsing matter cloud and 
$-\Lambda$ is the cosmological constant. Our solutions to the Wheeler-DeWitt 
equation are exact, so we are able to reliably compute the greybody factors that 
result from going beyond the near horizon region.

\medskip


%%%%%%%%%%%%%%%%%%%%%%%%%%%%%%%%%%%%%%%%%%%%%%%%%%%%%%%%%

\section{Introduction\label{intro}}

A principal goal of modern theoretical physics is the construction of a consistent
quantum theory of gravity. In fact much effort has been directed at this problem 
over several decades with the result that there are at present several proposals 
on the table. However, it is probably fair to say that none of the proposals 
currently under development is either fully understood, free of ambiguities or 
universally accepted \cite{OUP}.

In the absence of either a full quantum theory of gravity or of any direct 
experimental input, it seems worthwhile 
to address the quantization of particular
models by the application of as wide a variety of techniques as possible. Among the 
most interesting and best understood models are those with spherical
symmetry, while 
an especially interesting model with spherical symmetry is that of self-gravitating 
matter. One model of gravitational collapse that is very well understood on the 
classical level is that of Lema\^itre \cite{lemaitre}, Tolman \cite{tolman}, and Bondi 
\cite{bondi} (LTB), which describes a self-gravitating cloud of pressureless dust. 
The system is known to develop both covered and naked singularities and therefore a
successful quantization of these models has the potential to address also the fate of 
the naked singularity.

Ideally, one would like to have a quantum model that is able to predict what the final 
state of collapse will be, in particular whether or not and under what conditions the 
formation of a singularity may be avoided \cite{OUP,loopqg}. The model should also 
explain some features of the end-states. For example, it has been known for a very 
long time from semi-classical arguments that black holes behave as heat reservoirs 
and evaporate as black bodies with a characteristic temperature that depends only 
upon their conserved charges \cite{jdb,sh75,rw01}. Explaining these thermodynamic 
properties in terms of their microstates would be a desirable feature of any quantum 
theory of collapse. Further, semi-classical arguments indicate that the
nature of particle creation from a naked singularity is significantly different from 
a black hole and possibly even that they are unstable \cite{ns}; however, because the 
semi-classical treatment breaks down when curvatures reach the Planck scale, a full 
quantum treatment is essential to understanding their evolution. This is another 
area upon which a good quantum model of collapse should shed some light. 

In order to examine these problems we have set up a canonical quantization program 
\cite{vw99,vws01} to describe a self-gravitating dust cloud using 
the variables introduced in \cite{kk94}. In 
our work, the classical geometrodynamic constraints of the gravity-dust system were
given in terms of a canonical chart consisting of the mass contained within spherical
shells, the area radius, the dust proper time, and their conjugate momenta. The 
diffeomorphism constraint was used to eliminate the momentum conjugate to the mass 
function and this procedure was shown to result in a much simpler constraint that was 
able to take the place of the original Hamiltonian constraint. Dirac's quantization 
then led to the Wheeler--DeWitt equation for the wave functional describing the 
quantum collapse. Later it was shown that the WKB treatment of the Schwarzschild 
black hole in this canonical picture describes Hawking radiation \cite{vksw03}. To go
beyond the WKB approximation, the Wheeler--DeWitt equation was then regularized on a 
lattice and quantum corrections to the radiation were proposed in \cite{vws04}. However,
the lattice regularization turned out not to be differomorphism invariant. 
This issue was addressed in \cite{kmv06}. When care is taken to ensure that the momentum 
constraint is fulfilled in the continuum limit, the lattice wave functional becomes 
described by the Hamilton--Jacobi equation and two additional constraints which uniquely 
determine the factor ordering and make it possible to obtain exact solutions for the 
wave functional. Subsequently, we reexamined the Hawking radiation for the general 
case of non-marginal collapse \cite{kmsv07}, recovering the Planck spectrum near the 
horizon and additional grey-body factors as the near-horizon approximation was relaxed.

In this paper we turn our attention to the situation in which a negative cosmological
constant is present. We feel it is desirable to first adapt our program to the lower
dimensional 2+1-dimensional case, where dust collapse has been examined in detail on the classical 
\cite{roma, sg05} and the semi-classical level \cite{tpg07} by two of us. Moreover, the BTZ 
black hole \cite{btz92,bhtz93}, which is the unique classical end state of the collapse, is 
reasonably well understood on the quantum level from the point of view of the AdS/CFT 
correspondence \cite{as98,carlip}. Thus it may be possible in the future to compare the 
degrees of freedom employed by the two approaches. Most of the
calculations are in close analogy to the calculations presented in our
earlier papers on the LTB model, but there are various subtle differences
(boundary conditions, relative factors in expressions) that need to be
addressed; therefore, we present the analysis in sufficient detail to
provide a self-contained presentation.

Our paper is organized as follows. In Sec. II we review the classical model of 
self-gravitating dust in 2+1 dimensions and set up the canonical formalism in ADM 
variables. In Sec. III we transform to the more transparent Kucha\v r-type variables appropriate 
to the presence of a negative cosmological constant. Here we determine the mass function 
in terms of the ADM variables and go on to express the canonical constraints in a chart 
consisting of the radius, $R$, the mass function, $F$, the dust proper time, $\tau$ and 
their conjugate momenta. We also take special care to address the boundary action. In 
Sec. IV we obtain the Wheeler--DeWitt equation and present the exact quantum states 
appropriate to a lattice regularization. In Sec. V we show how Hawking radiation, 
together with the appropriate grey-body factors, can be recovered from our exact solutions. 
We summarize our results in Sec. VI and conclude with a few comments on potential 
future developments.

\section{The model}

\subsection{The classical solutions}

We are concerned with a self-gravitating, pressureless dust cloud, described by the 
energy-momentum tensor ${T^\mu}_\nu = \varepsilon~ U^\mu U_\nu$, in 2+1 dimensions 
with a negative cosmological constant $-\Lambda$, $\Lambda >0$.
The metric may be given in comoving, synchronous 
coordinates as
\beq
\label{metric1}
ds^2 = -d\tau^2 + e^{2b(\tau,\rho)} d\rho^2 + R(\tau,\rho)^2
d\varphi^2\ ,
\eeq
where $\tau$ is the dust proper time, and $\rho$ labels dust shells of physical (curvature) 
radius $R(\tau,\rho)$. Inserting this line element into Einstein's equations 
leads to the relation \cite{sg05}
\beq
e^{2b(\tau,\rho)}=\frac{(\partial_\rho R)^2}{2(E-F)},
\label{eineq1}
\eeq
where $E(\rho)$ and $F(\rho)$ are time independent but otherwise 
arbitrary functions of $\rho$, which satisfy
\beq 2\pi G
\varepsilon(\tau,\rho) = \frac{\partial_\rho F}{R(\partial_\rho R)},~~ (\partial_\tau R)^2 =
2E-\Lambda R^2.
\label{eineq}
\eeq
We note that the gravitational constant $G$ has the physical dimension of an inverse mass 
 in 2+1 dimensions (we set $c=1$ throughout); the quantities $F$ and $E$ are dimensionless.

The case of collapse is described by the condition $\partial_\tau R < 0$. There is still 
a freedom to rescale the shell label $\rho$: this can be fixed by demanding that
\beq
R(0,\rho) = \rho,
\eeq
so that at the initial epoch ($\tau=0$) the label coordinate is equal to the curvature 
radius $R$. This allows us to express the functions $E(\rho)$ and $F(\rho)$ in terms 
of the energy density at $\tau=0$. From (\ref{eineq}) we find
\bea
F(\rho) &=& 2\pi G\int_0^\rho \rho' \varepsilon(0,\rho') d\rho'\cr\cr
E(\rho) &=& [\partial_\tau R(0,\rho)]^2 + \Lambda \rho^2
\eea
The physical interpretation of these relations is that $2F(\rho)$ represents the 
gravitating mass inside the shell labeled by $\rho$, and $E(\rho)/2$
is the total energy per unit mass
of that shell. Therefore $F(\rho)$ is generally known as the ``mass function'' of the 
shell and $E(\rho)$ as its ``energy function''.

The solution to (\ref{eineq}),
\beq
R(\tau,\rho) = \sqrt{\frac{2E}{\Lambda}} \sin\left(-\sqrt{\Lambda} \tau + \sin^{-1}
\sqrt{\frac{\Lambda}{2E}}~ \rho\right),
\eeq
shows that the shell labeled by $\rho$ reaches a curvature radius $R=0$ at the time
\beq
\tau_0(\rho) = \frac 1{\sqrt{\Lambda}}\sin^{-1}\left(\sqrt{\frac{\Lambda}{2E}}~ 
\rho\right).
\label{sing}
\eeq
At $\tau=\tau(\rho)$ the shell becomes singular; therefore, $\tau$ can take values no 
larger than $\tau_0(\rho)$. A detailed analysis, performed in \cite{sg05}, shows that 
only those shells that obey the relation $2F>\Lambda R^2$ become trapped. Not only 
does this mean that, for a black hole to form, $F>0$ but also that a shell becomes 
trapped when it collapses to a size less than $\sqrt{2F/\Lambda}$.

The solutions can be matched to an exterior BTZ black hole \cite{roma, sg05},
\beq
ds^2 = -(\Lambda R^2 -GM) dT^2 + \frac{dR^2}{\Lambda R^2 -GM} + R^2 d\varphi^2,
\eeq
at some boundary that is specified by a fixed shell (the outermost shell), labeled 
$\rho_b$. The matching requires the mass function at the boundary to be related to 
the mass parameter of the BTZ black hole according to $F(\rho_b)=GM/2$.

\subsection{The Hamiltonian}

The ADM metric with circular symmetry for a 2+1 dimensional system takes the form
\beq
ds^2 = -N(t,r) dt^2 + L^2(t,r)(dt+N^r dr)^2 +R^2(t,r)d\varphi^2,
\label{ADM}
\eeq
where $N$ is the lapse function and $N^r$ is the only shift that survives the symmetry.
In terms of these metric components, the Einstein-Hilbert action can
be written as
\beq
S_\text{EH} = G^{-1}
\int dt \int_0^\infty dr\left[\frac{N}{L^2}(L'R'-LR''+\Lambda L^3R) -
\frac 1{N}[(N^r L)'- \dot L][N^r R' - \dot R]\right],
\eeq
after integrating out the angular dependence. From here, we obtain the momenta 
conjugate to $R(t,r)$ and $L(t,r)$,
\bea
P_L &=& -\frac 1 {GN} [\dot R - N^r R'],\cr\cr
P_R &=& -\frac 1 {GN} [\dot L - (N^r L)'],
\label{momenta}
\eea
and the action can be put in the form
\beq
S_{\text{EH}} = \int dt  \int_0^\infty dr [P_L \dot L + P_R \dot R - N \cH^g - N^r 
\cH_r] + S_{\partial \Sigma},
\label{action}
\eeq
where $S_{\partial \Sigma}$ is a boundary term to be discussed below. Because no time 
derivatives of the lapse or shift occur in the action, they are Lagrange multipliers. 
The Hamiltonian and momentum (diffeomorphism) constraints are given respectively by
\bea
\cH^g &=& -GP_L P_R - G^{-1}\Lambda RL +G^{-1}\left(\frac{R'}L\right)' \approx 0\cr\cr
\cH_r &=& R'P_R-LP_L' \approx 0.
\label{const1}
\eea
The total action is the sum of (\ref{action}) and an action $S_\text{D}$ that describes 
the dust. The canonical formalism for dust was developed in \cite{kt91} and elaborated 
in \cite{kb95}. We consider only non-rotating dust, for which
\beq
S_\text{D} = \int dt \int_0^\infty dr [P_\tau \dot \tau - N \cH^d -
N^r 
{\mathcal H}^d_r],
\eeq
where the Hamiltonian and momentum constraints are given by
\bea
\cH^d &=& P_\tau \sqrt{1+\frac{\tau'^2}{L^2}},\cr\cr
\cH^d_r &=& \tau' P_\tau
\eea
The matter configuration will also act as a time-keeper for the quantum theory. In 
principle one could think to use a fundamental field for this purpose, but this would 
make the problem much less tractable and, as we will see, the main features of the 
theory are already contained in the dust model.

It is easy to verify that the Poisson-bracket algebra of the constraints closes 
and the system is first class. With the total Hamiltonian denoted by $\bbH$,
the Hamiltonian equations of motion read
\bea
&& \dot R = \{R,\bbH\}_\text{PB} = \frac{\delta \bbH}{\delta P_R} = -NGP_L + N^r R',\cr\cr
&& \dot L = \{L,\bbH\}_\text{PB} = \frac{\delta \bbH}{\delta P_L} = -NGP_R + (N^r L)',\cr\cr
&& \dot \tau = \{\tau,\bbH\}_\text{PB} = \frac{\delta \bbH}{\delta P_\tau} = N\sqrt{1+
\frac{\tau'^2}{L^2}}+ N^r \tau',\cr\cr
&& \dot P_R = \{P_R, \bbH\}_\text{PB} =  -\frac{\delta \bbH}{\delta R} = -\frac{N''}{GL} +
\frac{N'L'}{GL^2} + G^{-1}N \Lambda L + (N^r P_R)',\cr\cr
&& \dot P_L = \{P_L, \bbH\}_\text{PB} =  -\frac{\delta \bbH}{\delta L} = -\frac{N' R'}{GL^2}
+ G^{-1}N \Lambda R + \frac{N\tau'^2 P_\tau}{L^2\sqrt{L^2+\tau'^2}}+N^r P_L',\cr\cr
&& \dot P_\tau = \{P_\tau, \bbH\}_\text{PB} =  -\frac{\delta \bbH}{\delta \tau} =
\left(\frac{N\tau' P_\tau}{L\sqrt{L^2+\tau'^2}} + N^r P_\tau\right)',
\eea
and from them the equations in (\ref{eineq}) can be recovered in the gauge $\tau = t$.

\subsection{The fall-off conditions}

We consider only mass functions which are such that at infinity the solutions 
approach the BTZ spacetime. This would be true in models in which the collapsing metric 
either asymptotically approaches or is smoothly matched to an exterior BTZ black hole 
at some boundary $\rho_b$. It is then necessary to adopt the following fall-off 
conditions at spatial infinity:
\bea
&& R(t,r) \rightarrow  r + \cO^\infty(r^{-2})\cr\cr
&& L(t,r) \rightarrow \frac{r^{-1}}{\Lambda^{1/2}} +\frac{GM_+(t)r^{-3}}{\Lambda^{3/2}} +
\cO^\infty(r^{-4})\cr\cr
&& P_R(t,r) \rightarrow \cO^\infty(r^{-4})\cr\cr
&& P_L(t,r) \rightarrow \cO^\infty(r^{-2})\cr\cr
&& N(t,r) \rightarrow \left[\sqrt{\Lambda}~ r -\frac{GM_+(t) r^{-1}}{\sqrt{\Lambda}} +
\cO^\infty(r^{-3})\right] N_+(t) + \cO^\infty(r^{-4})\cr\cr
&& N^r(t,r) \rightarrow \cO^\infty(r^{-2})
\eea
On the other hand, the fall-off conditions at the origin depend sensitively on the 
conditions we place on the energy and mass functions. If we wish to avoid shell 
crossing singularities then the energy function must be positive, with negative slope. 
Near the center ($r=0$) we take $E = \sum_n E_n r^n$, with $E_0 >0$ and $E_1 <0$. Likewise, 
take $F=\sum_n F_n r^n$ with $F_0>0$. In this way, we determine the following 
conditions as $r \rightarrow 0$:
\bea
&& R(t,r) \rightarrow a(t)+b(t)r + \cO^0(r^3),\cr\cr
&& L(t,r) \rightarrow \gamma b(t) + \cO^0(r),\cr\cr
&& P_R(t,r) \rightarrow P_{R0}(t) + \cO^0(r),\cr\cr
&& P_L(t,r) \rightarrow P_{L0}(t) + \cO^0(r),\cr\cr
&& N(t,r) \rightarrow \gamma N_0(t) + \cO^0(r^2),\cr\cr
&& N^r(t,r) \rightarrow \cO^0(r).
\eea
They are consistent with the constraints (\ref{const1}) and are preserved
by the equations of motion. With these fall-off conditions, the only non-vanishing 
boundary variations that arise read
\beq
\int dt N_+(t) M_+(t),
\eeq
where $N_+(t)$ is the lapse function as $r\rightarrow \infty$ and $GM_+=2F(r\rightarrow 
\infty)$ is the ADM mass and a similar contribution at $r=0$,
\beq
-\int dt N_0 (t) M_0(t)
\eeq
where $N_0(t)$ is the lapse function as $r\rightarrow 0$ and $GM_0=2F(0)$. To avoid the
conclusion that $N_+$ and $N_0$ freeze the evolution at the respective boundaries, the
boundary terms must be cancelled by an appropriate boundary action. This can be achieved
by adding the surface action
\beq
S_{\partial\Sigma} = -\int dt N_+(t) M_+(t)+ \int dt N_0(t) M_0(t)
\eeq
In the next section we show how this surface action can be absorbed into the hypersurface
action by introducing a new canonical chart.

\section{Canonical transformations}

\subsection{Mass function in terms of the canonical variables}

Now consider an embedding of the ADM metric (\ref{ADM}) in the
spacetime described by the metric (\ref{metric1}),
\beq
ds^2 = -d\tau^2+\frac{\tR^2}{2(E-F)} d\rho^2 +R^2 d\varphi^2,
\eeq
where $R=R(\tau,\rho)$, $\tR = \partial_\rho R$, and we will use $R^*=\partial_\tau R$ 
as opposed to a prime and a dot for derivatives with respect to the ADM labels $r$ 
and $t$, respectively. We shall set $G=1$ in the following to simplify
the expressions.
Let $\oR = \tR/\sqrt{2(E-F)}$, then
\bea
\label{L2NNr}
&& L^2 = \oR^2\rho'^2 -\tau'^2,\cr\cr
&& N = \frac{\oR}L (\dot \tau \rho' - \dot \rho \tau'),\cr\cr\
&& N^r = \frac{\oR^2 \dot \rho \rho' - \dot \tau \tau'}{L^2}.
\eea
Inserting these into the momenta (\ref{momenta}), we find that
\beq
LP_L = -\frac 1{\oR(\dot\tau \rho'-\dot \rho \tau')} \left[\dot R (\oR^2 \rho'^2 -
\tau'^2)-(\oR^2 \dot \rho \rho' - \dot\tau \tau')R'\right].
\eeq
Derivatives with respect to the ADM time can be exchanged for derivatives with respect 
to the proper time using
\bea
\dot R &=& R^* \dot \tau + \tR \dot\rho = R^* \dot\tau + \oR\sqrt{2(E-F)}~ 
\dot\rho,\cr\cr
R' &=& R^* \tau' + \tR \rho' = R^* \tau' + \oR \sqrt{2(E-F)}~ \rho',
\eea
which gives
\beq
LP_L = -\frac{R'R^*}{\sqrt{2(E-F)}}+\frac{(R^{*2} -2(E-F)}{\sqrt{2(E-F)}}\tau'.
\eeq
Substituting Einstein's equation (\ref{eineq}), $R^{*2}=2E-\Lambda R^2$, then yields 
the simplified form
\beq
LP_L = \mp\frac{R'\sqrt{2E-\Lambda R^2}}{\sqrt{2(E-F)}} - \frac{\Lambda R^2 -2F}
{\sqrt{2(E-F)}}\tau',
\eeq
which may be solved for $\tau'$:
\beq
\tau' =   -\frac{LP_L\sqrt{2(E-F)}}{\Lambda R^2 -2F} \mp \frac{R'\sqrt{2E-\Lambda R^2}}
{\Lambda R^2 -2F}.
\label{tauprime}
\eeq
Substituting this into the expression for $L$ in (\ref{L2NNr}),
\beq
L^2 = \oR^2 \rho'^2 - \tau'^2,
\eeq
and solving for $F$ gives an expression for the mass function in terms of the canonical
variables,
\beq
F=\frac 12\left[P_L^2-\frac{R'^2}{L^2}+\Lambda R^2\right].
\eeq
For future reference we introduce the function $\cF$ defined by
\beq
\cF = \Lambda R^2 - 2F = \frac{R'^2}{L^2}-P_L^2.
\eeq
$\cF=0$ determines the apparent horizon and $\cF$ will play an important role in the
quantum theory. One can check that though $\cF$ appears in the denominator of 
(\ref{tauprime}), $\tau'$ continues to be well behaved across the
horizon, as expected. (We note the difference in the definition of
${\mathcal F}$ compared to the LTB case where one has ${\mathcal
  F}=1-F/R$ \cite{kmv06}.)

\subsection{New variables}

As in the case of the Schwarzschild black hole \cite{kk94} 
and the LTB collapse model in 3+1 
dimensions \cite{vw99}, one can make a canonical transformation that elevates the mass 
function to a canonical variable. Interestingly, the expressions are similar to those 
presented by us in earlier papers on LTB collapse \cite{kmv06, kmsv07}.
By directly taking 
Poisson brackets the momentum conjugate to the mass function is found to be
\beq
P_F = \frac{LP_L}{\Lambda R^2-2F} =  \frac{LP_L}\cF.
\eeq
Looking for a canonical transformation that would take the diffeomorphism constraint to
\beq
\cH_r = R'P_R - LP_L' \rightarrow R'\oP_R + F' P_F,
\eeq
we find a simple expression for $\oP_R$:
\beq
\oP_R = P_R + \frac{\Lambda R L P_L}{\cF} - \frac{\Delta}{L^2 \cF},
\label{newmom}
\eeq
where $\Delta =  (LP_L)'R'-(LP_L)R''$. Our momentum constraint becomes
then indeed
\beq
\cH_r = R'\oP_R + F' P_F \approx 0,
\eeq
but we must first show that the transformation from the old set $\{R,L,P_R,P_L\}$ to the
set $\{R,F,\oP_R,P_F\}$ is canonical. We can do this by explicitly constructing the 
generator of the transformation. Denote it by $\bbF$. Since we already know two 
coordinates ($R$ and $F$) and one conjugate momentum ($P_F$), we use
\bea
P_L(r) &=& \int dr' P_F(r') \frac{\partial F(r')}{\partial L(r)} + \frac{\delta
\bbF}{\delta L(r)},\cr\cr
P_R(r) &=& \oP_R(r)+\int dr' P_F(r')\frac{\partial F(r')}{\partial R(r)} + \frac{\delta
\bbF}{\delta R(r)},\cr\cr
0 &=& \int dr' P_F(r') \frac{\partial F(r')}{\partial P_L(r)} + \frac{\delta \bbF}
{\delta P_L(r)},\cr\cr
0 &=& \int dr' P_F(r') \frac{\partial F(r')}{\partial P_R(r)} + \frac{\delta \bbF}
{\delta P_R(r)}.\cr
&&
\label{canonical}
\eea
The last equation in (\ref{canonical}) tells us that $\bbF=\bbF[R,L,P_L]$. The third 
equation gives 
\beq
P_F(P_L) + \frac{\delta \bbF}{\delta P_L} = 0 \Rightarrow \frac{\delta \bbF}{\delta P_L}
= \frac{LP_L^2}{P_L^2 - R'^2/L^2},
\eeq
and therefore
\beq
\bbF = \int dr \left[LP_L - R' \tanh^{-1} \frac{LP_L}{R'}\right] + \bbF_1[L,R],
\label{generating}
\eeq
whereas, from the first equation,
\beq
P_L + \frac{R'^2P_L}{L^2P_L^2 - R'^2} = \frac{\delta \bbF}{\delta L}
= P_L +\frac{R'^2 P_L}{L^2 P_L^2 - R'^2} + \frac{\delta \bbF_1}{\delta L},
\eeq
showing that $\bbF_1=\bbF_1[R]$. Take $\bbF_1$ to be independent of $R$, a constant,
and let us calculate $\oP_R$ from the resulting $\bbF$:
\beq
\oP_R = P_R -\int dr' P_F(r') \frac{\partial F(r')}{\partial R(r)}- \frac{\delta \bbF}
{\delta R}
\eeq
Integrating by parts,
\bea
\oP_R &=& P_R - \left(\frac{P_F R'}{L^2}\right)' + P_F \Lambda R - \frac{\delta \bbF}
{\delta R}\cr\cr
&=& P_R +\frac{\Lambda RLP_L}{\cF} - \frac{(LP_L/R')'}{1-(LP_L/R')^2}\cr\cr
&=& P_R + \frac{\Lambda RLP_L}{\cF} - \frac{\Delta}{L^2\cF},
\eea
we find precisely the new candidate momentum in (\ref{newmom}). The transformation
from $\{R,L,P_R,P_L\}$ to $\{R,F,\oP_R,P_F\}$ is generated by $\bbF$.

We now want to write the Hamiltonian in terms of the new variables, $\{R,\oP_R,F,P_F\}$.
To do so, we use the two equations
\bea
P_L^2-\frac{R'^2}{L^2} &=& -\cF,\cr\cr
LP_L &=&  \cF P_F.
\eea
Inserting the second into the first gives
\beq
\frac{P_F^2\cF^2-R'^2}{L^2}=-\cF\Rightarrow L^2= \frac{R'^2-P_F^2\cF^2}\cF
\eeq
and therefore
\beq
P_L^2= \frac{P_F^2\cF^3}{R'^2-P_F^2\cF^2}
\eeq
as well as
\beq
P_R = \oP_R +\Lambda RP_F -\frac{\Delta}{P_F^2\cF^2-R'^2}.
\eeq
Putting it all together, we find
\beq
\label{newHg}
\cH^g = -\frac 1L \left[\cF P_F\oP_R + \cF^{-1} R'F'\right]
\eeq
which is of the same form as the expressions for the Schwarzschild black hole and LTB
collapse in 3+1 dimensions, cf. Equation (42) in \cite{kmv06}. 
The action in the new canonical variables then reads
\beq
S_\text{EH} = \int dt \int_0^\infty dr \left(P_\tau \dot \tau+ \oP_R \dot R + P_F \dot F
- N \cH^g - N^r \cH_r\right) + S_{\partial \Sigma}
\eeq
with the new constraints (\ref{newHg}) and
\beq
\cH_r = R'\oP_R + F' P_F.
\eeq
We shall now discuss the boundary action in more detail.

\subsection{Boundary action}

The following considerations are in analogy to Sec.~II.D in \cite{kmv06}.
Because varying $N_+$ would lead to zero ADM mass and varying $N_0$ would restrict $F_0$
to zero, both $N_+$ and $N_0$ should be considered as prescribed functions. By the 
fall-off conditions, the lapse function, $N^r$, is required to vanish both at the center 
as well as at infinity. This implies that the time evolution is generated along the 
world lines of observers with $r=$ constant. If we introduce the proper time of these 
observers as a new variable, we can express the lapse function in the form $N_+(t) = 
\dot \otau_+$ and $N_0(t) = \dot \otau_0$. This leads to
\beq
S_{\partial\Sigma} = -\int dt M_+(t)\dot\otau_+ + \int dt M_0(t)\dot \otau_0.
\eeq
Thus we remove the need to fix the lapse function at the boundaries. Extending the 
treatment in \cite{kk94}, the aim is to cast the homogeneous part of the action into 
Liouville form, finding a transformation to new variables that absorb the boundary terms. 
This can be done by introducing the mass density $\Gamma = F'$ as a
new canonical variable. Define
\beq
F(r) = M_0 + \int_0^r dr'\Gamma(r),~~ \Gamma(r)=F'(r),
\eeq
and reconsider the Liouville form
\bea
\Theta &:=& \int_0^\infty dr P_F \delta F - M_+ \delta \otau_+ + M_0 \delta\otau_0\cr\cr
&=& \int_0^\infty P_F \delta F + \otau_+ \delta M_+ - \otau_0 \delta M_0,
\eea
where we have dropped an exact form. But
\beq
\delta F = \delta M_0 + \int_0^r dr' \delta \Gamma (r')
\eeq
gives
\beq
\Theta = \left(\int_0^\infty dr' P_F(r') - \otau_0\right)\delta M_0 + \int_0^\infty dr 
P_F(r) \int_0^r dr' \delta \Gamma(r') + \otau_+ \delta M_+.
\eeq
Noting further that\footnote{See Kucha\v r \cite{kk94}. Consider
\bean
&&\left(\int_0^r dr' \delta \Gamma(r') \times \int_{r}^\infty dr' P_F(r')\right)'\cr\cr
= && \delta \Gamma(r) \times \int_r^\infty dr' P_F(r')-P_F(r)\times\int_0^r dr'
\delta \Gamma(r')
\eea
Integrating the left hand side from $0$ to $\infty$ gives zero, therefore
\beqn
\int _{0}^\infty dr  P_F(r) \int_0^r dr' \delta \Gamma(r')=\int_{0}^\infty
dr~ \delta\Gamma(r)\int_r^\infty dr' P_F(r')
\eeq
}
\beq
\int _{0}^\infty dr  P_F(r) \int_0^r dr' \delta \Gamma(r')=\int_{0}^\infty
dr~ \delta\Gamma(r)\int_r^\infty dr' P_F(r'),
\eeq
we can write the Liouville form as
\bea
\Theta &=& \left(\int_0^\infty dr' P_F(r') - \otau_0\right)\delta M_0\cr\cr
&&\hskip 1cm  + \int_0^\infty dr \delta \Gamma(r) \left(\int_0^\infty dr P_F(r) -
\int_0^r dr' P_F(r')\right)+ \otau_+ \delta M_+ \cr\cr
&=& \left(\int_0^\infty dr' P_F(r') - \otau_0\right)\delta M_0\cr\cr
&&\hskip 1cm  + (\delta M_+-\delta M_0)\int_0^\infty dr P_F(r) - \int_0^\infty dr 
\delta \Gamma(r) \int_0^r dr' P_F(r')+ \otau_+ \delta M_+ \cr\cr
&=&p_0\delta M_0 + p_+ \delta M_+ + \int_0^\infty P_\Gamma(r) \delta \Gamma(r),
\eea
where
\bea
&& p_0 = -\otau_0,\cr\cr
&&p_+ = \otau_+ + \int_0^\infty dr P_F(r),\cr\cr
&&P_\Gamma(r) = - \int_0^r dr' P_F(r').
\label{congamma}
\eea
The new form of the action is then
\beq
S_\text{EH} = \int dt \left(p_0 \dot M_0 + p_+ \dot M_+ + \int dr~ [P_\tau 
\dot \tau+ \oP_R\dot R + P_\Gamma\dot \Gamma - N \cH^g - N^r \cH_r] \right),
\eeq
where the new constraints read
\bea
&&\cH^g = -\frac 1L\left[-\cF P_\Gamma'\oP_R + \cF^{-1} R'\Gamma \right] + 
P_\tau \sqrt{1+ \frac{\tau'^2}{L^2}}\approx 0,\cr\cr
&&\cH_r = R' \oP_R -\Gamma P_\Gamma' + \tau' P_\tau\approx 0.
\label{const2}
\eea
The Hamiltonian constraint can be simplified if the momentum constraint is used to
eliminate $P_{\Gamma}$, and the constraints in (\ref{const2}) can be replaced by the 
following equivalent set
\bea
&&P_\tau^2+\cF \oP_R^2 - \frac{\Gamma^2}{\cF} \approx 0\cr\cr
&&\cH_r = \tau' P_\tau + R' \oP_R -\Gamma P_\Gamma'\approx 0.
\label{constf}
\eea
(Note that the Hamiltonian constraint has been `squared' in order to
arrive at the new form; it is therefore of dimension mass over length
squared.
Re-inserting $G$ would correspond to the substitution $\Gamma\to\Gamma/G$.)
This involves a little algebra which has been described in the appendix of 
\cite{kmv06}. These equations will be used for quantization in 
the next section. 

Here we emphasize that the relative sign between the dust and gravitational 
kinetic terms can change because $\cF$ is greater than zero outside the horizon 
and less than zero inside. This change of sign is already present in (\ref{const2}) 
and has not been introduced by using the momentum constraint to eliminate $P_F$. 
It is of fundamental interest in the quantum theory because the Wheeler--DeWitt 
equation becomes locally elliptic outside and hyperbolic inside the horizon, 
which is of importance for the formulation of the proper boundary value problem. 
A change of sign was also noted for the case of a non-minimally coupled
scalar field \cite{k89}, see also \cite{bk97} for an earlier
discussion in the black-hole case.

\subsection{Relation between dust proper time and Killing time}

In what follows, we will have need of the relationship between the dust proper time 
and Killing time; therefore we now address this issue. 
This subsection corresponds to subsection II.E in the LTB case
\cite{kmv06}. We had obtained the expression 
for $\tau'$ in terms of the canonical variables in Sec. IIA as
\beq
\tau' = P_\Gamma'\sqrt{2(E-F)} \mp \frac{R'}\cF \sqrt{2(E-F)-\cF}
\eeq
Defining $a=1/\sqrt{2(E-F)}$ gives
\beq
\tau' = \frac{P_\Gamma'}{a} \mp \frac{R'}{a\cF} \sqrt{1-a^2\cF}.
\eeq
(The equation of motion guarantees that the quantity $\sqrt{1-a^2\cF}$ is real.) 
For a constant value of $a$, this equation can be integrated. When 
$R>\sqrt{2F/\Lambda}$,
\bea
a\tau &=& P_\Gamma \mp \int \frac{dR}\cF \sqrt{1-a^2\cF}\cr\cr
&=& P_\Gamma \pm \left[\frac{a}{\sqrt{\Lambda}}\tan^{-1}\frac{aR\sqrt{\Lambda}}
{\sqrt{1-a^2\cF}} + \frac 1{\sqrt{2\Lambda F}} \tanh^{-1} \frac{\sqrt{1-a^2\cF}}
{R\sqrt{\Lambda/2F}}\right]
\label{ptKT0}
\eea
and
\beq
a\tau = P_\Gamma \pm \left[\frac{a}{\sqrt{\Lambda}}\tan^{-1}\frac{aR\sqrt{\Lambda}}
{\sqrt{1-a^2\cF}} + \frac 1{\sqrt{2\Lambda F}} \tanh^{-1} \frac{R\sqrt{\Lambda/2F}}
{\sqrt{1-a^2\cF}}\right]
\label{ptKT}
\eeq
when $R<\sqrt{2F/\Lambda}$. We know that the spacetime surrounding a collapsing cloud 
in 2+1 dimensions with negative cosmological constant
is the BTZ black hole. The choice of sign depends on whether one 
is interested in an expanding (positive sign) or collapsing (negative sign) cloud. 
Matching the collapse and BTZ metrics also shows that $T=P_\Gamma$ at the boundary.

Equations \eqref{ptKT0} and \eqref{ptKT} may still be used so long as $E'$ and $F'$ 
are sufficiently small, since then we have a small amount of dust propagating in the 
BTZ background. In that case, it would give the relationship between the time used 
by families of freely falling observers and the Killing time, each family being 
characterized by a fixed value of $E$.

\subsection{Interpretation of the canonical data}

We have already reconstructed the mass function from the canonical data. In this
section we want to reconstruct the energy function $E$ and singularity curve $\tau_0$
from the same. The three functions $E$, $F$, and $\tau_0$ determine the collapse model
completely, so reconstructing these quantities gives physical meaning to the
canonical variables.

If we begin with the momentum constraint in the form
\beq
\tau' = -\frac{R'\oP_R}{P_\tau} + \frac{\Gamma P_\Gamma'}{P_\tau}
\eeq
and use the Hamiltonian constraint to eliminate $\oP_R$, we get
\beq
\tau' = \pm \frac{R'}{\cF}\sqrt{\frac{\Gamma^2}{P_\tau^2}
+\cF} + \frac{\Gamma P_\Gamma'}{P_\tau}.
\eeq
Substituting $P_\Gamma' = - P_F = LP_L/\cF$, we find an expression,
\beq
\tau' = \pm \frac{R'}{\cF}\sqrt{\frac{\Gamma^2}{P_\tau^2}
+\cF} + \frac{\Gamma L P_L}{\cF P_\tau}
\eeq
that may be directly compared with the expression we had in (\ref{tauprime}) for
$\tau'$. We see that
\beq
P_\tau = \frac{\Gamma}{\sqrt{2(E-F)}},
\label{ptau}
\eeq
and therefore the energy function is related to the canonical variables by
\beq
E = \frac{\Gamma^2}{2(\Gamma^2/\cF - \cF \oP_R^2)} + F
\eeq
where we have used the Hamiltonian constraint to give the result in terms of
gravitational phase space variables only. Finally, knowing $E$ and $F$ in terms of
the canonical variables, it is a simple matter to do the same for the singularity
curve using the solution (\ref{sing}), cf. \cite{kmv06}.

\subsection{Hamilton equations of motion}

We now give the Hamilton equations of motion for the new system and derive Einstein's
equations (\ref{eineq}) from them. Introducing the smeared constraints
\bea
&&H[N] = \int_0^\infty dr N(r) \cH(r),\cr\cr
&&H_r[N^r] = \int_0^\infty dr N^r(r), \cH_r(r)
\eea
and the Hamiltonian $\bbH=\cH[N] + \cH_r[N^r]$, the canonical equations of the 
evolution are\footnote{Note that
\beqn
\frac{\delta F(r')}{\delta \Gamma(r)} = \Theta(r'-r),
\eeq
where $\Theta$ is the step function.}
\bea
&&\dot M_0 = \frac{\delta \bbH}{\delta p_0} = 0,\cr
&&\dot p_0 = -\frac{\delta \bbH}{\delta M_0} = 2\int_0^\infty dr~ N \left(\oP_R^2
+ \frac{\Gamma^2}{\cF^2}\right),\cr
&&\dot M_+ = \frac{\delta \bbH}{\delta p_+} = 0,\cr
&&\dot p_+ = -\frac{\delta \bbH}{\delta M_+} = 0,\cr
&&\dot \tau = \frac{\delta \bbH}{\delta P_\tau} = 2 N P_\tau + N^r \tau',\cr
&&\dot P_\tau = -\frac{\delta \bbH}{\delta \tau} =  (N^rP_\tau)',\cr
&&\dot R = \frac{\delta \bbH}{\delta \oP_R} = -2 N \cF \oP_R +N^r R',\cr
&&\dot \oP_R = -\frac{\delta \bbH}{\delta R} =  2N \left(\Lambda R \oP_R^2 +
\frac{\Lambda R \Gamma^2}{\cF^2} \right) + (N^r \oP_R)',\cr
&&\dot \Gamma = \frac{\delta \bbH}{\delta P_\Gamma} = (N^r \Gamma)',\cr
&&\dot P_\Gamma = -\frac{\delta \bbH}{\delta \Gamma} = \frac{2N\Gamma}
{\cF},\cr
&&\hskip 2cm  + 2 \int_0^r dr' N(r') \left[\oP_R(r')^2 + \frac{\Gamma(r')^2}
{\cF(r')^2}\right]+ N^r P_\Gamma'.
\eea
To obtain Einstein's equations, we notice that because
\beq
R^* = \frac{\dot R}{\dot \tau} \stackrel{N^r=0}{=} -\frac{\cF\oP_R\sqrt{2(E-F)}}{\Gamma},
\eeq
the momentum $\oP_R$ conjugate to $R$ may be expressed as
\beq
\oP_R = -\frac{\Gamma R^*}{\cF\sqrt{2(E-F)}}.
\eeq
Substituting this result in the Hamiltonian constraint leads directly to (\ref{eineq}).
Note that we did not have to specify the lapse function.

The lapse no longer has the interpretation of being the ratio between the proper 
time to BTZ time. The reason is that we have squared the original version of the 
Hamiltonian constraint. If we define a new version of the constraint by taking the 
square root,
\beq
\cH_\uparrow = P_\tau - \sqrt{-\cF \oP_R^2 +\frac{\Gamma^2}\cF},
\eeq
then
\beq
\{\tau,\cH_\uparrow[N^\uparrow]\} = N^\uparrow
\eeq
shows that it is $N^\uparrow$ that carries that interpretation.

Finally, we remark that the algebra of the constraints cannot be of the general form
(given for example in \cite{OUP}), again because we have used the momentum constraint 
to eliminate $P_F$ in the Hamiltonian constraint. In fact, a short calculation gives
\bea
&&\left\{\mathcal{H}[N], \, \mathcal{H}[M]\right\} = 0,\\
&&\left\{\cH_r[N^r], \, \cH[N]\right\} = \cH[N_{,r} N^r - N N_{,r}^r],\\
&&\left\{\cH_r[N^r], \, \cH_r[M^r]\right\} = \cH_r\left[[N^r,M^r]\right].
\eea
The Poisson bracket of the Hamiltonian with itself vanishes,
in contrast to the general case where it closes on the momentum constraint. The
other brackets coincide with the general case. The transformations generated
by the Hamiltonian constraint can thus no longer be interpreted as hypersurface
deformations. They are in general not orthogonal to the hypersurfaces, but act
along the dust flow lines.

\section{Exact quantum states}

\subsection{Quantization and ansatz}

We now apply Dirac's quantization procedure to turn the classical constraints
into quantum operators
\beq
P_X = -i \frac{\delta}{\delta X(r)},
\label{diracq}
\eeq
and acting the operators on wave-functionals. The Hamiltonian constraint then
reads
\beq
\left[\frac{\delta^2}{\delta \tau^2} + \cF \frac{\delta^2}{\delta R^2} + A\delta(0)
\frac{\delta}{\delta R} + B\delta(0)^2-\frac{\Gamma^2}{\cF}\right] \Psi[\tau,R,
\Gamma] = 0.
\label{qham}
\eeq
where $A$ and $B$ are smooth functions of $R$ and $F$ that encapsulate the factor
ordering ambiguities. The factor ordering problem is unsolved and can be dealt with
only after a suitable regularization procedure is implemented. We have
included
the formal expression
$\delta(0)$ to indicate the need for this regularization. Quantizing the momentum constraint
using (\ref{diracq}) gives
\beq
\left[\tau'\frac{\delta}{\delta \tau} + R'\frac{\delta}{\delta R} - \Gamma
\left(\frac{\delta}{\delta \Gamma}\right)' \right] \Psi[\tau,R,\Gamma] = 0,
\label{qdiff}
\eeq
but, as noted, the quantum constraints in (\ref{qham}) and (\ref{qdiff}) are only 
formal until a regularization procedure has been selected.

To regularize, we follow the construction \cite{kmv06} and consider a one dimensional 
lattice of discrete points $r_i$ separated by a distance $\sigma$ which must be later 
taken to zero to achieve the continuum limit. We will require the wave-functional to 
(i) automatically satisfy the momentum constraint in the continuum limit and (ii) be 
factorizable into different functions for each lattice point. Such a
wave functional is of the form 
\bea
\Psi[\tau,R,\Gamma] &=& \exp\left[i\int dr \Gamma(r) \cW(\tau(r),R(r),F(r))\right]\cr\cr
&=&\lim_{\sigma\rightarrow 0} \prod_i \exp\left[i\sigma \Gamma_i \cW_i(\tau_i,R_i,F_i)
\right],
\eea
where $F_i = \sum_{j=0}^i \sigma \Gamma_i$. Details of the construction are given 
in \cite{kmv06}. One finds that the Hamiltonian constraint is satisfied independently 
of the choice of $\sigma$ only if the following three equations are simultaneously 
obeyed:
\bea
&&\left[\left(\frac{\partial \cW_j}{\partial\tau_j}\right)^2 + \cF_j 
\left(\frac{\partial \cW_j}{\partial R_j}\right)^2 - \frac{1}{\cF_j}\right]=0,\cr\cr
&&\left[\frac{\partial^2 \cW_j}{\partial\tau_j^2} + \cF_j\frac{\partial^2 \cW_j}
{\partial R_j^2} + A_j \frac{\partial \cW_j}{\partial R_j} \right]  =  0,\cr\cr
&& B_j = 0.
\label{3eqns}
\eea
The first equation is the Hamilton--Jacobi equation and was used in \cite{vksw03}. The 
second equation presents an additional restriction on solutions and the last equation 
tells us that working on the lattice is only possible if the factor ordering does not 
contribute to the potential term. If we find solutions to all three equations, we can 
do all other calculations on the lattice, since these solutions have a well defined 
continuum limit and satisfy the momentum constraint. We will now find exact solutions 
to all three equations.

\subsection{The measure}

The function $A$ is closely tied to the inner product on the Hilbert space via the 
hermiticity of the operator
\beq
\widehat{\cF_j P_{j,R}^2} = \cF_j \frac{\partial^2}{\partial R_j^2} + A_j \frac{\partial}
{\partial R_j}.
\eeq
It is therefore intimately connected to the choice of measure, $\mu_j$. Below
we consider the relationship between $\mu$ and $A$, assuming that $\mu$,
is independent of $\tau$. For $\hO=\widehat{\cF P_R^2}$ to be hermitian, we require
\beq
\int_0^\infty dR \mu(R) \phi^*(R) \{\hO\psi(R)\} = \int_0^\infty dR \mu(R)\{\hO
\phi(R)\}^* \psi(R),
\eeq
which leads, provided the boundary conditions are trivial, to the (coupled) system
\bea
&&A_j^* = -A_j + \frac{2}{\mu_j}\frac{\partial}{\partial R_j}(\mu_j\cF_j),\cr\cr
&&\frac{\partial^2}{\partial R_j^2}(\mu_j \cF_j) - \frac{\partial}{\partial R_j}(\mu_j
A_j)=0.
\eea
The second equation can be integrated to give
\beq
\frac{1}{\mu_j}\frac{\partial}{\partial R_j}(\mu_j\cF_j) - \frac{\cQ_j(F_j)}{\mu_j} = A_j,
\label{measure}
\eeq
and the first requires that $\Re e(\cQ)=0$. We can view (\ref{measure}) as a (Bernoulli) 
equation for $\mu_j$:
\beq
\frac{\partial \mu_j}{\partial R_j} + \frac{1}{\cF_j}\left(\frac{\partial \cF_j}
{\partial R_j}- A_j\right)\mu_j = \frac{\cQ_j(F_j)}{\cF_j},
\label{mueqn}
\eeq
whose general solution is
\beq
\mu_j = e^{-\int \cP_j(R_j)dR_j}\left[\pm \cQ_j(F_j)\int dR_j \frac{e^{\int\cP_j
(R_j)dR_j}}{|\cF_j|}+ \alpha_j\right],
\label{musol}
\eeq
where the upper sign is for the exterior ($\cF>0$), the lower sign is for the interior 
and
\beq
\cP_j=\frac{1}{|\cF_j|}\left(\frac{\partial |\cF_j|}{\partial R_j}\mp A_j \right),
\eeq
again with the same sign conventions.

For solutions of (\ref{3eqns}), discussed in the following subsection, the relation 
between $A$ and $\mu$,
\beq
A_j= |\cF_j| \partial_{R_j} \ln (\mu_j|\cF_j|),
\eeq
which re-expresses (\ref{measure}) with $\cQ_j(F_j)=0$ (because $\cQ_j(F_j)$ can 
only be imaginary), will be of greater interest than (\ref{musol}).

\subsection{Exact solutions}

We have already emphasized that the signature in the kinetic part of the Hamiltonian
constraint can change from elliptic (outside the horizon) to hyperbolic (inside the
horizon). This carries over to the kinetic term of the Wheeler-DeWitt equation.
As discussed in \cite{bk97}, we can say that the part inside the horizon is always
classically allowed, whereas this is not necessarily the case for the outside part. 
The usual initial value problem appropriate for hyperbolic equations can thus only 
be applied for the region corresponding to the black hole interior.

We determine the unique, exact solutions to (\ref{3eqns}) in Appendix A and here 
we summarize the results. Both in the exterior, $\cF>0$, and interior, $\cF<0$, we 
obtain a two parameter family of solutions: for the measure
\beq
\mu_j = \frac{\beta_j}{\sqrt{1-a_j^2\cF_j}},
\label{meas}
\eeq
and for $\cW_j$
\beq
\cW_j = \text{const.} \pm a_j \tau_j \pm \int dR_j \frac{\sqrt{1-a_j^2\cF_j}}{\cF_j},
\label{Wj}
\eeq
where $a_j$ and $\beta_j$ are constants of integration. The physical significance 
of $a_j$ can be demonstrated by acting on our wave functional with the dust energy 
operator $\hP_\tau$: 
\beq
\hP_\tau \Psi_a = a \Gamma \Psi_a = \frac{\Gamma}{\sqrt{2(E-F)}} \Psi_a,
\eeq
showing that $a=1/\sqrt{2(E-F)}$. The integral appearing in (\ref{Wj}) will be 
recognized then as the integral we evaluated earlier in (\ref{ptKT}) in order to 
obtain the relationship between the dust proper time and the BTZ time. The 
wave functionals are therefore oscillatory in the classically allowed regions.

Given the requirement of factorizability on the lattice, there are no
other solutions (either separating or non-separating) that solve the constraints, 
that is, we have obtained the complete class of factorizable solutions to all 
the constraints. Other solutions would necessarily couple the shells composing the 
collapsing cloud. The solutions we have obtained, which are strictly valid on 
the lattice, contain two free parameters, $a_j$ and $b_j$, and we write them as
\beq
\Psi_j = e^{-\sigma b_j \Gamma_j } e^{\pm i \sigma \Gamma_j (a_j \tau_j \pm \int dR_j
\frac{\sqrt{1-a_j^2\cF_j}}{\cF_j})}.
\eeq
We know that $a_j$ is connected with the energy and mass functions and that in 
general both $E$ and $F$ are functions of the radial label coordinate. It would 
therefore be natural to demand that $a$ is also an arbitrary function of $r$, but 
this explicit dependence on $r$ would violate the momentum constraint. The way out 
is to require the $r$ dependence of $a$ to appear via the mass function, $F(r)$, 
{\it i.e.,} we must require that $a=a(F(r))$, and likewise $b=b(F(r))$. Any 
dependence on $r$ via the mass function is allowed by the momentum constraint and 
since the Hamiltonian constraint does not contain a derivative with respect to 
$\Gamma$ or $F$ it continues to be obeyed. Hence, in the continuum limit we arrive at
\beq
\Psi[\tau,R,\Gamma] = e^{-\int_0^\infty dr \Gamma(r) b(F(r))}\exp\left\{\pm i 
\int_0^\infty dr \Gamma(r)  \left[ a(F(r))\tau \pm \int dR \frac{\sqrt{1-a^2(F(r))\cF}}
{\cF}\right]\right\}.
\label{exst}
\eeq
The fact that $a(r)$ is constrained by diffeomorphism invariance to depend on 
$r$ only via $F(r)$ means that the energy function is not arbitrary in the quantum 
theory but only those energy functions expressible in terms of $F$, {\it i.e.,} 
$E(r) = F(r) + 1/2a^2(F(r))$ are allowed.

\section{Hawking radiation}

\subsection{Introduction}

The exact quantum states in (\ref{exst}) describe the generic situation. In order
to describe Hawking radiation, we need to introduce into the formalism the black 
hole and the analogue of the quantum fields which are used in the standard treatment.

Following \cite{vksw03} and \cite{kmsv07}, we consider a BTZ black hole surrounded 
by tenuous dust, by taking the mass function to be of the form $2F(r) = GM\Theta(r) 
+ 2f(r)$, 
where $\Theta(r)$ is the Heaviside function and $2f(r)$ represents the 
dust perturbation.  Formally, the variable $\Gamma=GM\delta(r)+f'(r)$ describes 
this perturbation and $f'(r)$ describes the mass density outside $r=0$. As described 
in \cite{vksw03}, the black hole state factors into a product state, one member 
of the which represents the black hole, while the other represents the dust
and describes the Hawking radiation.

The central idea of \cite{vksw03} was to use the Bogoliubov transformation of the 
field operators in the BTZ spacetime. Here, since we have the exact quantum states 
at our disposal, we first identify the states that correspond to the ingoing and 
outgoing modes, respectively, of the standard approach and then calculate their inner 
product. The inner product is evaluated on hypersurfaces of constant Killing time, 
which correspond to observers who are static with respect to the black hole, not 
constant dust time, which would correspond to observers who are freely falling. We 
will therefore express our solutions in terms of the BTZ time, $T$, using (\ref{ptKT}).

The relevant exact solutions of the Wheeler--DeWitt equation read (in
this section we re-introduce $\hbar$ and $G$)
\beq
\Psi^{\pm}=\exp\left(-\frac 1{G\hbar}\int dr b \Gamma
\mp\frac{i}{2G\hbar}\int dr \ \Gamma\left[a\tau\pm \int^R dR\
    \frac{\sqrt{1-a^2{\mathcal F}}}{\mathcal F}\right]\right)\ .
\label{N2}
\eeq
We will not concern ourselves with the normalizations at this point because they 
make no difference to our final result.

For the positive frequency, incoming wave functional we choose $\Psi^+$ from
(\ref{N2}) and replace the dust proper time with the Killing time using (\ref{ptKT}).
We find
\beq
\label{in}
\Psi^+=\exp\left(-\frac 1{G\hbar}\int dr b \Gamma
 -\frac{i}{G\hbar}\int dr\ \Gamma T\right)\ ,
\eeq
which has the standard form of a positive-frequency wave function. This wave-functional
is independent of $R$. The lattice version of this state is obtained through the 
replacement of $\Gamma=F'$ by a frequency $\omega$, and 
is given by
\beq
\label{in2}
\Psi^+_{\sigma\omega}=\lim_{\sigma\to 0} \prod_j
e^{(-\sigma b\omega_j -i\sigma\omega_j T_j)/G\hbar}\ .
\eeq
For the out-going modes of negative frequency  we have to take the state $\Psi^-$. 
Inserting (\ref{ptKT}) into the corresponding state of (\ref{N2}) then gives
\beq
\Psi^-=\exp\left(-\frac 1{G\hbar}\int dr b \Gamma +
\frac{i}{G\hbar}\int dr\ \Gamma\left[T-2\int^R dR\
    \frac{\sqrt{1-a^2{\mathcal F}}}{\mathcal F}\right]\right)\ ,
\eeq
the corresponding lattice version being
\beq
\label{out}
\Psi^-_{\sigma\omega}=\lim_{\sigma\to 0}\prod_j
\exp\left(-\frac {\sigma b_j \omega_j }{G\hbar} +
\frac{i\sigma\omega_j}{G\hbar}\left[ T_j-2\int dR_j\
    \frac{\sqrt{1-a_j^2{\mathcal F}_j}}{\mathcal F_j}\right]\right)\ .
\eeq

\subsection{The Bogoliubov coefficient in the near horizon limit}

The Bogoliubov coefficients \cite{bd84,kmsv07} are given by
\beq
\label{N6}
\beta_{\omega\omega'} =
\frac{2\sigma\omega}{G\hbar} \int_{R_h}^{\infty}dR\
\sqrt{g_{RR}}\Psi^{-*}_{\sigma\omega}\Psi^+_{\sigma\omega'} \ .
\eeq
Using the coordinate transformation from $(R,\tau)$ to $(R,T)$, we have in $(R,T)$ 
coordinates,
\beq
\label{grr}
\sqrt{g_{RR}}=\frac{1}{a(\Lambda R^2-2F)}\ ,
\eeq
where $a=\frac{1}{\sqrt{2(E-F)}}$. Inserting the wave functionals (\ref{in2}) and
\eqref{out} into \eqref{N6} then gives
\bea
&& \beta_{\omega\omega'}= \frac{2\omega\sigma}{G\hbar} \sqrt{2(E-F)}
\exp\left(-\frac{\sigma b(\omega+\omega')}{G\hbar} -i\frac{\sigma
  T(\omega+\omega')}{G\hbar}\right)\times \nonumber\\
&& \int_F^\infty dR \frac{1}{\Lambda R^2-2F} \exp\left[
\frac{2i\sigma\omega}{G\hbar}
\int^R dR \frac{\sqrt{1-a^2{\mathcal
        F}}}{\mathcal{F}}\right]\ .
\eea
We now change the variable of integration to $\cF$, keeping in mind the near horizon
approximation. $\cF$ goes to zero on the horizon and diverges as $R$ goes to infinity.
We have ${d\cF}/{2\Lambda R}=dR$. Rewriting the integral in the new
variable yields
\bea
&&\beta_{\omega\omega'} = \frac{2\omega\sigma}{G\hbar} \sqrt{2(E-F)}
\exp\left(-\frac{\sigma b(\omega+\omega')}{G\hbar} -i\frac{\sigma
  T(\omega+\omega')}{G\hbar}\right)
\times\cr\cr
&& \int_0^\infty d\cF \frac{1}{2\cF \sqrt{\Lambda (\cF+2F)}}
 \exp\left[\frac{2i \sigma\omega }{G\hbar} \left(\int^{\cF} d\tilde{\cF}
 \frac{\sqrt{(1 -a^2\tilde{\cF})}}
{2\tilde{\cF} \sqrt{\Lambda (\tilde{\cF}+2F)}}\right)\right].
\label{exact}
\eea
In the near horizon limit, we expand about $\cF=0$. The integral becomes
\bea
\beta_{\omega\omega'} &=& \frac{\sigma\omega}{G\hbar \sqrt{2\Lambda F}}\sqrt{2(E-F)}
\exp\left(-\frac{\sigma b(\omega+\omega')}{G\hbar} -i\frac{\sigma
  T(\omega+\omega')}{G\hbar}\right)
\times \cr\cr
&& \int_0^\infty d\cF\ \cF^{-1+\frac{i\sigma\omega }{G\hbar \sqrt{2 \Lambda F}}}
\exp\left[-i\frac{\sigma\omega }{G\hbar \sqrt{2 \Lambda F} }
\left(\frac{a^2}{2}+\frac{1}{4F}\right)\cF\right]\ .
\label{N10}
\eea
Using the following formula after inserting the regularization factor $e^{-p\cF}$ for convergence,
\begin{displaymath}
\int_0^{\infty}dx\ x^{\nu-1}e^{-(p+iq)x}=\Gamma(\nu)(p^2+q^2)^{-\nu/2}
  e^{-i\nu\mathrm{arctan}(q/p)},
\end{displaymath}
where $\nu=\frac{i\sigma\omega }{G\hbar \sqrt{2 \Lambda F}}$ and $q=\frac{\sigma\omega }
{G\hbar \sqrt{2 \Lambda F} } \left(\frac{a^2}{2}+\frac{1}{4F}\right)\cF $, we find
\bea \label{N10a}
& & \beta_{\omega\omega'}=\frac{\sigma\omega}{G\hbar \sqrt{2 \Lambda F}}\sqrt{2(E-F)}
\exp\left(-\frac{\sigma b(\omega+\omega')}{G\hbar} -i\frac{\sigma
  T(\omega+\omega')}{G\hbar}\right)
\times \nonumber\\
& & \ \Gamma\left(\frac{i\sigma\omega }{G\hbar \sqrt{2 \Lambda F}}\right)
   \left(\frac{\sigma\omega }{G\hbar \sqrt{2 \Lambda F}} \left(\frac{a^2}{2}+\frac{1}{4F}\right)\right)^{-i\sigma\omega
/(G\hbar \sqrt{2 \Lambda F})}e^{-\pi\sigma\omega /(2G\hbar \sqrt{2 \Lambda F})}\ .
\eea
The absolute square of the above expression is given by
\beq\label{N10b}
\vert\beta_{\omega\omega'}\vert^2 = \frac{2\pi
  \sigma\omega(2(E-F))}{G\hbar \sqrt{2 \Lambda F}}
\frac{e^{-2\sigma b(\omega+\omega')/G\hbar}}{e^{2\pi \sigma\omega
    /G\hbar \sqrt{2 \Lambda F}} - 1}\
\eeq
and determines the particle creation via
\beq
\langle \text{in}| {\hat N}_\text{out}|\text{in}\rangle = 
\int_0^{\infty}d(\sigma\omega')\
\vert\beta_{\omega \omega'} \vert^2.
\eeq
(We integrate here over $\sigma\omega'$ in order to obtain a
 dimensionless expression.)
Performing the integration and replacing 
$\sigma \omega$ by $G\Delta \epsilon$, where $\Delta\epsilon$ denotes
the energy of a shell, we get (setting also $2F=GM$)
\beq
\label{N11}
\langle\mathrm{in}\vert\hat{N}_{\mathrm{out}}\vert\mathrm{in}\rangle=\frac{\pi
  \Delta \epsilon(2E-GM))}{b \sqrt{ \Lambda GM}}
\frac{e^{-2b\Delta \epsilon/\hbar}}{e^{2\pi \Delta \epsilon
    /\hbar \sqrt{ \Lambda GM}} - 1}\ ,
\eeq
where ${\hat N}_\text{out}$ represents the ``out'' particle number.

We recognize that the Planck spectrum is modified by greybody factors which 
explicitly depend on the energy. 
These greybody factors are different from those 
obtained by taking into account back-scattering (for example, see \cite{Page}) 
because we are using exact solutions of the quantum constraints. Both the 
gravitational field and the dust are quantized in our approach.

The constant $b$ that appears in the final expression for $\langle\text{in}| 
N_\text{out} |\text{in}\rangle$ also has its origin in the full quantum 
gravitational state. The simplest choice for $b$ is $b=0$, in which case the 
integral over $\omega'$ diverges. This divergence, however, is well-known and is connected with the normalization of 
the continuous modes. Dividing out 	the infinite constant from this integration 
the result, for $b=0$, is\footnote{We divide by $G\hbar$ in order to
  keep the result dimensionless.}
\beq
\langle\mathrm{in}\vert\hat{N}_{\mathrm{out}}\vert\mathrm{in}\rangle
=\frac{\pi
  \Delta \epsilon(2E-GM))}{\hbar \sqrt{ \Lambda GM}}\frac{1}{e^{2\pi 
\Delta \epsilon
    /\hbar \sqrt{ \Lambda GM}} - 1}\
\eeq
This is a thermal
spectrum with temperature given by
\beq
\label{temperature}
k_{\mathrm{B}}T_{\mathrm{H}}=\hbar \frac{\sqrt{\Lambda GM}}{2\pi} \ ,
\eeq
which holds for each shell separately.

\subsection{Exact Bogoliubov coefficient}

It is possible to obtain an exact expression for the Bogoliubov coefficient as well by
considering a series expansion about the horizon. We expand equation (\ref{exact})
around $\cF=0$ to all orders. Expanding the measure of equation (\ref{exact}), we get,
\beq
\label{measure1}
\frac{\sqrt{1+\frac{\cF}{2F}}}{2\sqrt{\Lambda 2F}\cF}=\frac{1}{2\sqrt{\Lambda 
2F}}\frac{1}{\cF}\left(1-\frac{\cF}{4F}+\frac{3}{8}\left(\frac{\cF}{2F}
\right)^2\ldots\right) \stackrel{\rm{def}}{=} \sum_{n=-1}^\infty \alpha_n s^n,
\eeq
Similarly the exponent can be rewritten as,
\bea
&&\frac{i\sigma\omega }{G\hbar \sqrt{2 \Lambda F}}\left( \ln{\cF} -
\left(\frac{a^2}{2}+\frac{1}{4F}\right)\cF+  ....
\right)\cr\cr
\stackrel{\rm{def}}{=}
 &&\frac{i\sigma\omega }{G\hbar \sqrt{2\Lambda F}}
\left(\ln \cF + \sum_{m=1}^\infty \beta_m \cF^m\right)\ ,
\eea
The integral can be performed before summing over the series of (\ref{measure1}).
\bea
&&\frac{\sigma\omega}{G\hbar \sqrt{2\Lambda F}}\sqrt{2(E-F)}~ \times\cr\cr
&&\hskip 1cm \sum_{n=-1}^\infty \alpha_n\int_0^\infty d\cF\
\cF^{n+i\sigma\omega /(G\hbar \sqrt{2\Lambda F})}e^{i\sigma\omega \beta_1 
\cF/(G\hbar \sqrt{2\Lambda F})} e^{i\sigma\omega \sum_{m=2}^\infty \beta_m 
\cF^m/(G\hbar\sqrt{2\Lambda F})}
\eea
where $\beta_1=-(1/a^2+1/4F)$.
 \bea
&&e^{i\frac{\sigma\omega}{(G\hbar \sqrt{2\Lambda F})} \sum_{m=2}^\infty \beta_m 
\cF^m} = 1+\frac{i\sigma\omega }{G\hbar \sqrt{2\Lambda F}}\sum_{m=2}^{\infty}
\beta_m\cF^m\cr\cr
&&\hskip 4.5cm +\frac 12\left(\frac{i\sigma\omega}{G\hbar \sqrt{2\Lambda F}}
\right)^2\sum_{m,n=2}^{\infty} \beta_m\beta_n\cF^m\cF^n\equiv \sum_{m=0}^\infty 
\gamma_m \cF^m,
\eea
where $\gamma_0=1$ and $\gamma_1=0$. The integral is
\bea
&&=\frac{\sigma\omega}{G\hbar \sqrt{2\Lambda F}}\sqrt{2(E-F)}\sum_{n,m=0}^\infty
\alpha_{n-1}\gamma_m \int_0^\infty d\cF \cF^{-1+n+m+i\sigma\omega /(G\hbar 
\sqrt{2\Lambda F} )}
e^{i\sigma\omega \beta_1 \cF/(G\hbar \sqrt{2\Lambda F})}\cr\cr
&&=\left.\frac{\sigma\omega}{G\hbar \sqrt{2\Lambda F}}\sqrt{2(E-F)}~\times \right.
\cr\cr
&&\hskip 2cm \left.\sum_{n,m=0}^\infty \alpha_{n-1}\gamma_m \left(-i\frac{\partial}
{\partial \chi}\right)^{n+m}\int_0^\infty
d\cF \cF^{-1+i\sigma\omega F/(G\hbar \sqrt{2\Lambda F})} e^{i\chi \cF}
\right|_{\chi=\sigma\omega \beta_1/(G\hbar \sqrt{2 \Lambda F})}.
\eea
and the exact Bogoliubov coefficient can be written as
\bea
\beta_{\omega\omega'} &=&\frac{\sigma\omega}{G\hbar \sqrt{2 \Lambda F}}\sqrt{2(E-F)}
\exp\left(-i\frac{\sigma T(\omega+\omega')}{G\hbar}\right)
  \times\cr\cr
&&\left(\frac{\sigma\omega \beta_1}{G\hbar \sqrt{2 \Lambda F}}\right)^{-i\sigma
\omega /(G\hbar \sqrt{2 \Lambda F})}e^{-\pi\sigma\omega /(2G\hbar \sqrt{2 
\Lambda F})}\Gamma \left(\frac{i\sigma\omega }{G\hbar \sqrt{2 \Lambda F}}\right)
\times \left[1 + \right.\cr\cr
&& \left.\sum_{n+m=1}^\infty (-i)^{n+m}\alpha_{n-1}\gamma_m \left(-\frac{i\sigma
\omega }{G\hbar \sqrt{2 \Lambda F}}\right)\times \ldots \times \left(-\frac{i\sigma
\omega }{G\hbar\sqrt{2 \Lambda F}}-n-m+1\right)\times \right.\cr\cr
&& \hskip 10cm \left. \left(\frac{\sigma\omega \beta_1}{G\hbar \sqrt{2 \Lambda 
F}}\right)^{-n-m}\right].\cr
&&
\label{fullbeta}
\eea
The simplest correction term to the near-horizon approximation is $\omega$-independent
and obtained for $n=1$, $m=0$:
\begin{displaymath}
(-i)\alpha_0\gamma_0\left(-\frac{i\sigma\omega
    }{G\hbar \sqrt{2 \Lambda F} }\right)\left(\frac{\sigma\omega
    \beta_1}{G\hbar \sqrt{2 \Lambda F}}\right)^{-1}=-\frac{\alpha_0}{\beta_1}\ .
\end{displaymath}
$\alpha_0=-1/4F$ and $\beta_1=-(a^2/2+1/4F)$, which can be simplified to be 
$-\frac{1}{2a^2E}$.

In general the higher order corrections are $\omega$ dependent and therefore lead to
non-trivial modifications of the greybody factors. However, in the limit as $\omega 
\rightarrow \infty$, the corrections once again become independent of $\omega$. 
These correction terms cannot be obtained in the standard derivation of the Hawking 
radiation because the geometric optics approximation is strictly assumed. We can 
obtain them because we have the exact quantum state at our disposal.

\section{Discussion}

In this paper we have extended the canonical formalism originally developed for the
Schwarzschild black hole and later for the LTB collapse of
inhomogeneous dust in 3+1 dimensions to 2+1 dimensional collapse with a negative
cosmological constant. In this work, our emphasis has been on the physical 
parameters $E$, $F$, and $\tau_0$, and we have succeeded in expressing them in 
terms of the canonical variables. This renders the formalism more transparent.

Using a lattice regularization scheme that correctly implements the diffeomorphism 
constraint in the continuum limit and assuming factorizability on the lattice, we 
presented exact and unique solutions to the quantum constraints and extended them 
to the continuum. They coincide with the WKB approximation, which therefore become 
exact. 

Using the canonical formulation we produced a general relationship between the dust 
proper time and the Killing time and then used our solutions to derive the Hawking 
radiation from a BTZ black hole. It was possible to reliably calculate the greybody 
factors because our solutions are exact. No further corrections to the Hawking radiation 
are obtained because the WKB solution is exact within the framework of our regularization.

It is worthwhile comparing our derivation of Hawking radiation with
the corresponding calculation in the four-dimensional LTB case
\cite{kmsv07}, and also with the semiclassical 2+1-d derivation of
Hawking radiation \cite{tpg07}.
There are significant differences 
in the gravitational collapse between the 2+1-dimensional and the 3+1-dimensional case on the classical level. The effects are indeed captured in the quantum formalism developed in the paper. Through canonical transformations both Wheeler--DeWitt equations can be transformed to the same form. The qualitative  difference between the two cases is brought out by the difference in the form of $\mathcal{F}$ which captures the information about the details of trapped surface formation. In the calculation of Hawking radiation, the framework is the same as in the four-dimensional case. In both cases it yields the correct Hawking temperature, though the correction terms for the Bolgoliubov coefficients turn out to be different.

In \cite{tpg07}, Hawking radiation is obtained semiclassically by
considering the quantization of scalar field modes in the background
of a BTZ spacetime. The modes are shown to have the asymptotic form
which resembles the standard plane wave form with a decay factor. The
behavior at spatial infinity had to be taken care of by using
reflecting boundary conditions since the AdS spatial infinity is
timelike. The scalar field is equated to zero at the spatial
infinity. In our case, we have the wave functionals for dust. In order
to evaluate the Bogoliubov coefficients, we assumed a scalar product
of the form given in (\ref{N6}). The integral is evaluated over the
spatial slice of constant Killing time. The value of the wave
functional for each shell  is asymptotically decaying for large
$R$. This makes the wave functional zero at spatial boundary of the
AdS. Thus our solution automatically addresses the problem of the AdS
spatial infinity being timelike.  

There are several outstanding issues that must remain for future work. One of particular
interest is to obtain a microscopic description of the entropy of the BTZ black hole.
The BTZ black hole is a special case of our solutions, corresponding
to a constant mass function. Its entropy was computed using the AdS/CFT correspondence, 
although the physical significance of the degrees of freedom being counted in that approach 
remains ambiguous. An advantage of the canonical approach is that the degrees of freedom 
have a transparent physical meaning. A comparison between the microscopic degrees of freedom 
from the canonical theory and those counted using the AdS/CFT would be illuminating. 

Another issue concerns the study of a possible singularity avoidance in the simplified 
realm of 2+1 dimensions. Singularity avoidance has been claimed from loop quantum 
gravity \cite{loopqg} and has been shown to hold for collapsing dust shells in 3+1 
dimensions \cite{hkmp01}. The quantum collapse of shells in 2+1 dimensions was recently 
examined in \cite{ortizryan07} and the results suggest that at least a portion
of the collapsing shell rebounds, reemerging from the horizon in a finite proper time, 
while the remaining portion of the shell is lost to Hawking radiation. 

A third issue concerns the role of naked singularities in the quantum theory. This problem 
is of particular experimental interest because semi-classical arguments suggest that 
naked singularities are quantum mechanically unstable. If those arguments hold, naked 
singularities could be experimental windows into the world of quantum gravity.

Finally, the fact that the WKB approximation becomes exact in the lattice regularization
means that it is possibly useful to look for alternative regularization schemes. We plan
to address some of these issues in future publications.

\bigskip

\noindent{\bf Acknowledgements} 

\noindent TPS gratefully acknowledges support from the
German Science Foundation (DFG) under the grant 446 IND 113/34/0-1.
\vfill\eject

%%%%%%%%%%%%%%%%%%%%%%%%%%%%%%%%%%%%%%%%%%%%%%%%%%%%%%%%%%%%%%%%%%%%%%%%%%%%%

\centerline{\bf APPENDIX : Uniqueness of solutions}

\bigskip

In this appendix we discuss the uniqueness of the solutions in \eqref{exst}. We
will consider the solutions in the exterior ($\cF>0$). The solutions in the interior
are obtained in an analogous manner by replacing the trigonometric substitution below
by the analogous hyperbolic one.

For $\cF>0$, the Hamilton-Jacobi equation [the first of \eqref{3eqns}] is solved by
taking
\bea
\frac{\partial \cW_j}{\partial \tau_j} &=& \frac{\cos\eta_j}{\sqrt{\cF_j}},\cr\cr
\frac{\partial \cW_j}{\partial R_j} &=& \frac{\sin\eta_j}{\cF_j}.
\label{constraints3}
\eea
Integrability of this solution implies that
\beq
\frac{\partial \eta_j}{\partial \tau_j} = -\sqrt{\cF_j}\tan\eta_j\frac{\partial
\eta_j}{\partial R_j}-\frac{\partial \sqrt{\cF_j}}{\partial R_j},
\label{integrability2}
\eeq
but, inserting (\ref{constraints3}) into the second equation of \eqref{3eqns}, we
find another equation for $\eta(\tau,R,F)$:
\beq
-\frac{\sin\eta_j}{\sqrt{\cF_j}}\frac{\partial \eta_j}{\partial \tau_j} + 
\cos\eta_j \frac{\partial \eta_j}{\partial R_j}-\sin\eta_j \frac{\partial \ln\cF_j}
{\partial R_j} + \frac{A_j}{\cF_j}\sin\eta_j =0,
\eeq
which, using the relationship between $A$ and the measure $\mu$ derived earlier, simplifies
to
\beq
\frac{\partial \eta_j}{\partial \tau_j}=\sqrt{\cF_j}\cot\eta_j \frac{\partial
\eta_j}{\partial R_j}+\sqrt{\cF_j}\frac{\partial \ln \mu_j}{\partial R_j}.
\label{second2}
\eeq
However, equations \eqref{second2} and \eqref{integrability2} are consistent if and 
only if
\beq
\tan\eta_j = \frac{\alpha_j}{\mu\sqrt{\cF_j}},
\eeq
where $\alpha_j=\alpha_j(\tau_j)$ is a function {\it only} of $\tau_j$. Taking 
derivatives with respect to $\tau_j$ and $R_j$ respectively and reinserting them into 
either the integrability condition (\ref{integrability2}) or into (\ref{second2}) now 
gives an equation for $\alpha(\tau)$,
\beq
\frac{\partial \alpha_j}{\partial \tau_j} =\frac{\alpha_j^2}{\mu_j^2} \frac{\partial\mu_j}
{\partial R_j} - \frac{\mu_j}{2} \frac{\partial \cF_j}{\partial R_j}.
\eeq
Now as $\alpha$ is a function of $\tau$ but not of $R$, and $\mu, \cF$ are functions of $R$ but
not of $\tau$, the above equation requires $\alpha = \text{const.}$ and therefore
\beq
\frac{\alpha_j^2}{\mu_j^3} \frac{\partial\mu_j}{\partial R_j} = \frac{1}{2}\frac{\partial \cF_j}
{\partial R_j}.
\eeq
Thus we obtain the solutions (cf. \eqref{meas} and \eqref{Wj})
\beq
\mu_j = \frac{\beta_j}{\sqrt{1 -a_j^2 \cF_j}},~~ \cW_j = \text{const.} \pm a_j\tau_j \pm
\int dR_j \frac{\sqrt{1-a_j^2\cF_j}}{\cF_j},
\eeq
where $a_j$ and $\beta_j=a_j\alpha_j$ are constants. They are unique under the
given conditions.

\bigskip

\bigskip

%%%%%%%%%%%%%%%%%%%%%%%%%%%%%%%%%%%%%%%%%%%%%%%%%%%%%%%%%%%%%%%%%%%%%%%%%%%%%%
   
%%%%%%%%%%%%%%%%%%%%%%%%%%%%%%%%%%%%%%%%%%%%%%%%%%%%%%%%%%%%%%%%%%%%%%%%%%%%%%%%%%%%%%%%%
\end{document}